\documentclass{jfp}

\usepackage{appendix}
\usepackage{graphicx}
\usepackage{ecltree}
\usepackage{epic}
\usepackage{moreverb}

\title{The Heap Lambda Machine}

\author[A.~Salikhmetov]{\uppercase{Anton Salikhmetov} \\
RTSoft, P.~O.~Box~158, 105077 Moscow, Russia \\
\email{salikhmetov@gmail.com}}

\begin{document}

\maketitle

\begin{abstract}

This paper introduces a new machine architecture for evaluating
lambda expressions using the normal-order reduction, which
guarantees that every lambda expression will be evaluated if the
expression has its normal form and the system has enough memory.
The architecture considered here operates using heap memory only.
Lambda expressions are represented as graphs, and all algorithms
used in the processing unit of this machine are non-recursive.

\end{abstract}

\section{Introduction}
\label{IntroSec}

Automated evaluation of lambda expressions has drawn attention of
many researchers.
A number of different approaches to design machines that directly
deal with lambda expressions has been proposed in the literature,
and the monograph~\cite{Kluge} gives a comprehensive overview of
many such designs.

We have noticed that all such machines relied upon quite
complicated memory structure and required rather intricate memory
management techniques.
Typically, the memory is subdivided into several functionally
different areas.
Among such areas can be stacks, environments, code areas, heaps,
and so on. 
Such arrangements imply the need to specify a separate interface
to each memory subsystem: a stack pointer register to keep track
of stack utilization, a dynamic memory allocator for heaps,
garbage collectors, etc.
Besides, conventional computer memory provides just a linear
array of identical memory cells, each cell being addressable by
its index in this array.
For such memory, it remains unclear as to which criteria should
be employed for partitioning the array into functionally
different parts.

These observations motivated us to investigate whether it is
possible to construct a machine possessing the following two
properties.
First, the memory should be uniform, i.e. no subdivision of the
former into functionally different parts such as a stack and a
dynamic memory area was allowed.
Second, we wanted the memory management mechanisms to be
super-simple, with their algorithmic implementation and the
interface being as minimal as possible.

It appears that it is indeed possible to satisfy the requirements
mentioned above.
Having started from the idea of graph reduction, we designed the
machine where the entire memory is a uniform collection of
sequentially addressable blocks allocated on demand.
We have also implemented a portable software emulator of this
machine.

The memory manager in our machine consists of a single register
and three commands only.
Taking into account the similarity of our memory allocator and
heap-based dynamic memory allocators, we have decided to refer
this machine to as the Heap Lambda Machine.
Worth mentioning here is also the fact that careful design of the
processing unit algorithms allowed us to avoid using garbage
collection.

The purpose of this paper is to describe the architecture of our
machine and to demonstrate all vital parts of the emulator.

\section{High-Level Design and the User's View of the System}
\label{DesignSec}

\begin{figure}[b]
\begin{center}
\includegraphics[height=90mm]{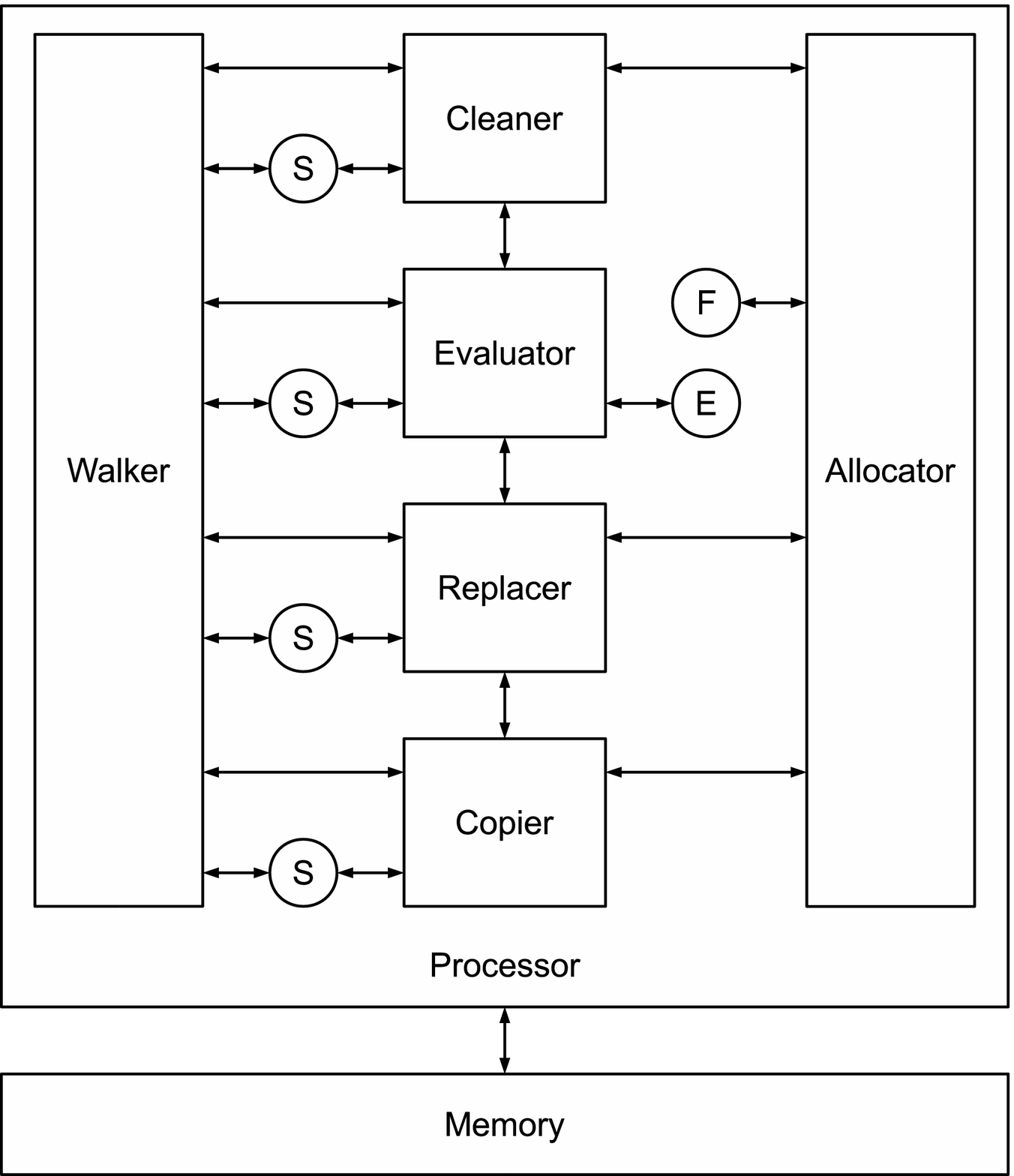}
\end{center}
\caption{The architecture of the Heap Lambda Machine.
In the block diagram, F denotes the \texttt{freehead} register,
E stands for the \texttt{expr} register, and the state registers
are shown with the S letter.}
\label{DesignFig}
\end{figure}

The system consists of several units shown in
\figurename~\ref{DesignFig}, the main units being the memory and
the processor.
The units can interact by transferring control and data as
indicated in the block diagram by the arrows.
In some cases, units use common data of the special \emph{state}
type explained in Section~\ref{WalkSec}.
Concrete structure of the units depends on a particular
implementation of the machine: in the abstract machine, these
are simply algorithms described in later sections of this paper;
in our software emulator the units are C language functions; had
this machine been implemented in hardware, each unit would have
been a microprogram using a set of internal registers and
communicating with its neighbors by asserting electrical signals.

The memory in our machine is externally visible, i.e. the user
can read and write to it.
The entity to govern the memory usage is the memory manager,
consisting of the Allocator and the \texttt{freehead} register.
The Allocator exports the interfaces to initialize the memory as
well as to allocate and free its units.
In the software emulator, the machine memory is modeled via an
array obtained using the Standard C Library function
\texttt{calloc}; in the abstract machine, the memory is an array
of identical sequentially addressable blocks.
The user has to prepare the lambda expression using the internal
format explained in Section~\ref{MemorySec}, allocate a
sufficient amount of machine memory, load the expression into
memory, load the memory address where the expression starts into
the \texttt{expr} register, and transfer control to the
Evaluator.

The Evaluator is the entry point to the machine.
When evaluation is over, the user can read the result from the
machine memory starting from the address in \texttt{expr} and
optionally convert it into a suitable format.
In our software emulator, the Evaluator is implemented as a C
function, so that when this function returns, this means to the
caller that evaluation is complete.
As for the abstract machine, we do not specify any particular
mechanisms to signal the end of the computation; if this machine
were implemented in hardware, such mechanisms would be defined
at the hardware design stage.

The  Walker, Cleaner, Replacer, and Copier units are helper
blocks in the processor, and these units are not intended to be
visible to the user.
Their design and implementation are described later on.

\section{The Memory Model}
\label{MemorySec}

In our machine, lambda expressions are represented as
graphs---this idea has become standard after~\cite{Wadsworth}.
The machine memory containing the lambda expression under
evaluation has linear structure and consists of blocks, each
block representing a single node of the lambda expression graph.

A node in the memory is a record of four address cells.
The first one called \texttt{par} points to the parent node.
The second one is called \texttt{copy} and is used during copying
of subexpressions as well as in order to link free blocks
(see Section~\ref{StorageSec} below).
The two remaining cells called \texttt{func} and \texttt{arg}
hold the addresses of the subexpressions, if any.
Additionally, their contents define the type of the node.

In the usual manner, we have three types of lambda expression
nodes: an application, an abstraction, and a variable.
In the case of an application, the \texttt{func} cell points to
the operator subexpression, while the \texttt{arg} cell points to
the operand subexpression---both \texttt{func} and \texttt{arg}
cells are non-zero.
For abstractions, the \texttt{func} cell points to the function
body subexpression, and \texttt{arg} contains the null pointer.
Finally, for variables, the \texttt{func} cell is zero,
\texttt{arg} points to an abstraction node which it is linked
with.

For example, the $apply$ combinator
$\lambda x. \lambda y. (x\, y)$ will be represented in the
machine memory as shown in \figurename~\ref{MemoryFig}.

\begin{figure}[ht]
\begin{center}\begin{tabular}{cccccc}
\hline\hline
Address & \multicolumn{4}{c}{Cell} & Expression \\
& \,\texttt{par}\, & \,\texttt{copy}\, & \,\texttt{func}\, &
\,\texttt{arg}\, & \\
\hline
1 & 0 & 0 & 2 & 0 & $\lambda x. \lambda y. (x\, y)$ \\
2 & 1 & 0 & 3 & 0 & $\lambda y. (x\, y)$ \\
3 & 2 & 0 & 4 & 5 & $(x\, y)$ \\
4 & 3 & 0 & 0 & 1 & $x$ \\
5 & 3 & 0 & 0 & 2 & $y$ \\
\multicolumn{6}{c}{\dotfill} \\
\hline\hline
\end{tabular}\end{center}
\caption{The memory dump for the $apply$ combinator.}
\label{MemoryFig}
\end{figure}

The well-known issue with name clashes \cite{Barendregt} for the
variable names is avoided in our machine automatically thanks to
the fact that different variables are represented as pointers to
different nodes.
Effectively, numeric block addresses in our memory model play the
role of variable names.

The free variable nodes are represented with null pointers in the
address cells, i.e. both \texttt{func} and \texttt{arg} are zero.
We think that it is convenient to treat the free memory blocks as
nodes that represent fictional free variables: indeed, such
blocks formally have the type of a free variable node.

\section{Storage Management}
\label{StorageSec}

Before the lambda expression can be processed, the machine memory
should be initialized as described below.
Initially, every memory block is put into the linked list of free
blocks similar to that discussed in Section 16.2 of~\cite{Field}.
Traversing from the last block till the first
one, the machine links them into the free nodes list using the
\texttt{copy} cell as the pointer to the next node.
The register called \texttt{freehead} is to hold the head node of
this list and points to address 1 at the beginning of the system
lifecycle.
The initial state of the machine memory is illustrated in
\figurename~\ref{InitFig}.
Our software emulator implements memory initialization via the
\texttt{reset} command shown in~\appendixname~\ref{StorageApp}.

\begin{figure}[b]
\begin{center}\begin{tabular}{cccccc}
\hline\hline
Address & \multicolumn{4}{c}{Cell} & Expression \\
& \,\texttt{par}\, & \,\texttt{copy}\, & \,\texttt{func}\, &
\,\texttt{arg}\, & \\
\hline
1 & 0 & 2 & 0 & 0 & $x$ \\
2 & 0 & 3 & 0 & 0 & $x$ \\
\multicolumn{6}{c}{\dotfill} \\
$N - 1$ & 0 & $N$ & 0 & 0 & $x$ \\
$N$ & 0 & 0 & 0 & 0 & $x$ \\
\hline\hline
\end{tabular}\end{center}
\caption{The initial state of the machine memory of size $N$
blocks, each block representing a fictional free variable.}
\label{InitFig}
\end{figure}

The machine allocates and frees nodes by manipulating the linked
list of free blocks and changing the contents of the
\texttt{freehead} register accordingly via the following two
commands: \texttt{get} and \texttt{put}.

If the \texttt{freehead} register contains a non-zero value, the
\texttt{get} command saves the node the \texttt{freehead}
register points to and updates this register by the value in the
\texttt{copy} cell of the saved node.
Then, \texttt{get} zeroes out each cell in the saved node and
returns it to the caller.
In the case when the \texttt{freehead} register contains the zero
value, which means that the system is out of free memory, calling
\texttt{get} triggers a machine exception and evalution is
aborted.

In turn, the \texttt{put} command takes one operand---the address
of the block to be put back into the free blocks list.
This command sets the \texttt{copy} cell of its operand to the
value kept in the \texttt{freehead} register, then changes the
latter to the address received as the operand.

For more details of the system initialization and storage
management, please see \appendixname~\ref{StorageApp}.

\section{Walking Through the Expression Tree}
\label{WalkSec}

Most of central mechanisms in the machine rely upon the ability
to traverse the tree in normal order, which in our case means
that the function part of an application is processed first.
The algorithm of tree traversing is factored out into a separate
unit, the fundamental idea behind this unit being that of a
state.
The state consists of the following three components: the current
node address, the address of the parent node of the subexpression
being traversed, and the direction (forth, i.e. towards the child
node, or back, i.e. towards the parent).
Based upon this state, a command called \texttt{walk} decides
which path should be followed at a particular step, makes this
step and returns the type of the step chosen: a
variable---direction is set to backward, a function part---the
current node is changed to the function part, an argument
part---direction is set to forward and the current node is
changed to the argument part, going back---the current node is
changed to the parent node, or finish---the state is not changed,
but the \texttt{walk} command indicates that walking is complete.
Note that this mechanism is a modification of the pointer
reversing approach explained in Section 11.3.2 of~\cite{Field}.
Note also that our walking algorithm is non-recursive, hence
using stacks is avoided.

Before walking through the expression tree, it is necessary to
initialize the state using a special command called
\texttt{init}.
The initial state has the direction forth, the current node
address pointing to the subexpression node, and the parent node
address pointing to the parent of the subexpression.
\appendixname~\ref{WalkApp} presents the implementation of this
unit.

\begin{figure}[b]
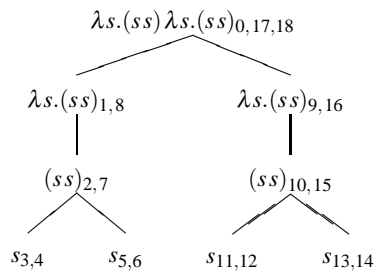

\begin{center}
\begin{bundle}{$\lambda s. (s\, s)\, \lambda s. (s\, s)_{0,\,
17, 18}$}
\chunk{\begin{bundle}{$\lambda s. (s\, s)_{1,\, 8}$}
	\chunk{\begin{bundle}{$(s\, s)_{2,\, 7}$}
		\chunk{$s_{3, 4}$}
		\chunk{$s_{5, 6}$}
	\end{bundle}}
\end{bundle}}
\chunk{\begin{bundle}{$\lambda s. (s\, s)_{9,\, 16}$}
	\chunk{\begin{bundle}{$(s\, s)_{10,\, 15}$}
		\chunk{$s_{11, 12}$}
		\chunk{$s_{13, 14}$}
	\end{bundle}}
\end{bundle}}
\end{bundle}
\end{center}
\caption{Traversal order for the tree representing the $\Omega$
combinator.}
\label{TreeFig}
\end{figure}

\figurename~\ref{TreeFig} shows an example of traversing through
the expression tree.
Here the node subscripts indicate the step numbers at which this
particular node is traversed.

\setlength{\GapWidth}{24pt}

\section{Clearing Subexpressions}
\label{ClearSec}

Clearing of subexpressions is needed after the replacement of
bound variables with respective subexpressions to put now useless
blocks back to the free blocks list.

Tree walking is the basic mechanism subexpression clearing is
based upon.
It can be easily seen that given the tree traversal strategy
described above, freeing the child nodes every time when the
walker has just gone up will necessarily result in freeing the
whole tree.
For instance, for the expression tree shown in
\figurename~\ref{TreeFig}, steps 7, 8, 15, 16, and 17 are the
places where the child nodes are freed.

For more details about implementation of the \texttt{clear}
command described in this Section, please see
\appendixname~\ref{StorageApp}.

\section{Copying Subexpressions}
\label{CopySec}

While replacing the bound variables with respective
subexpressions, i.e. with the argument part of an application
whose function part is an abstraction, the machine is copying
the argument subexpression using the command called
\texttt{copy}.
This command uses the walking mechanism as well as the
\texttt{clear} command described in Section~\ref{ClearSec}.

In contrast to \texttt{clear}, \texttt{copy} considers every
value the \texttt{walk} command returns in order to appropriately
construct a copy and move through the new expression being
constructed.
Construction itself is made on the steps of the following types:
an argument part, a function part, and an argument.
When going back, the pointer to the current node of a new
expression under construction is changed to its parent.
Each of the steps listed above was described in
Section~\ref{WalkSec}.

The most complicated problem within the \texttt{copy} command is
that variables in the new subexpression should point to the
corresponding abstractions.
Indeed, if the abstraction nodes are just constructed, the
command should map the pointer in the variable nodes from the one
in the original subexpression to those in the copy.
In the machine, this problem is solved as described below.

While walking through the original subexpression under copying,
two cases of \texttt{walk} steps are processed in a special
manner: a function part and a variable.

In the first case, the parent of the current node in the original
subexpression is changed: its \texttt{copy} cell is set to the
address of the corresponding node in the new subexpression.
Such way, the mapping of old abstractions to the new ones is
constructed.

In the case of a variable, the \texttt{copy} command searches for
the abstraction the original variable node points to by going
back through the whole expression.
When the corresponding abstraction node is found and its
\texttt{copy} cell contains a non-zero value, \texttt{copy} sets
the \texttt{arg} cell value to the value in the \texttt{copy}
cell of the found node.

The implementation of the \texttt{copy} command described above
can be found in \appendixname~\ref{CopyApp}.

\section{Replacing Bound Variables}
\label{ReplaceSec}

Evaluation of lambda expressions requires replacement of bound
variables in function bodies with the copies of arguments.
To make such a copy the machine uses the \texttt{copy} command
described in Section~\ref{CopySec}.
As to searching for bound variables in a function body, a special
command called \texttt{replace}, which walks the subexpression
tree and locates bound variables, is introduced.

The \texttt{replace} command takes three operands, each operand
representing a pointer to a subexpression node.
The first operand means the subexpression where the command
should look for the bound variable which corresponds to the
abstraction pointed to by the second one.
The third one contains the subexpression whose copy should be
substituted for the bound variable just found.
After substitution has finished, \texttt{replace} puts the bound
variable node back to the free blocks list using the \texttt{put}
command discussed in Section~\ref{StorageSec}.

For more details of replacement algorithm implementation, please
see \appendixname~\ref{ReplaceApp}.

\section{The Evaluation Algorithm}
\label{EvalSec}

In order to evaluate lambda expression in the memory, the machine
walks through the expression tree and looks for nodes that can be
reduced.
The reducibility check for a node is performed by a separate
command called \texttt{isreducible}, which returns a boolean
value at a subexpression node.
The \texttt{isreducible} command examines whether the node
represents an application.
If this is the case, it checks if the function part of the
application is an abstraction.
In the case when both conditions are satisfied, the command
returns true, otherwise it returns false.
Implementation of this command can be found in
\appendixname~\ref{ReplaceApp}.

When a reducible node is found, this node (which is the current
one from the viewpoint of the walker) is an application having
an abstraction in its operator part.
Using the \texttt{replace} command (Section~\ref{ReplaceSec}),
the machine makes one step of beta reduction.
When this step is complete, the application node, as well as the
abstraction node, ceases to exist as part of the expression.
Recall that \texttt{replace} makes copies of the argument for
each entry of the bound variable---that is, the entire
application operand subexpression is not needed anymore.
Hence the memory allocated for the application, the abstraction
and the operand can and should be freed.
This is the place where the \texttt{clear} command described in
Section~\ref{ClearSec} is used: note that in order to clear all
these entities properly it suffices to zero out the \texttt{func}
cell of the abstraction node and start clearing from the node
which represents the application.

When the current node represents an operator part of an
application, the algorithm changes the current node to the parent
because the latter may be now the leftmost outermost redex---such
behavior is the consequence of the fact that the machine makes
use of the normal-order reduction.

For more details about implementation of the \texttt{normal}
command described above, please see \appendixname~\ref{EvalApp}.

\section{Conclusions}
\label{ResumeSec}

This paper presented a detailed description of the machine for
automated evaluation of lambda calculus expressions.
Major features of this machine include using graphs to represent
lambda expressions, a memory manager of ultimate simplicity, and
normal order evaluation.
The uniform structure of the machine memory and the idea of ``the
entire memory is heap'' is what distinguishes our approach from
the ones previously found in the literature.

All algorithms of the processing unit were exposed in great
detail, and the concept of the machine has been proven by
implementing a portable software emulator; for the latter, this
paper includes the source code of all core parts of it in the
form of a C library.
In the simplest case, this library will be linked to an
application, which provides a human interface to the machine.
Please note that full sources of the machine emulator including
an implementation of the human interface are available as
Web-accessible accompanying material for this paper.

Our further research will concentrate on the following topics.
First, we will attempt to implement lazy
evaluation~\cite{Wadsworth}.
Second, we will explore the design of a more sophisticated I/O
model rather than using the entire memory for information
exchange between the machine and its outside world.
Of course, all above extensions of the Heap Lambda Machine are to
be done without sacrificing the simplicity of its memory
management.

\appendix

\section{The Library Interface}
\label{HeaderApp}

The following is the header file \texttt{machine.h} that
describes the library interface and contains declarations of all
needed data types, functions, and global variables.
Interesting to the library user are the \texttt{lambda} data
type, which represents a pointer to a node in the lambda
expression graph, the \texttt{get} function, which should be
called to allocate memory for a node, and the \texttt{normal}
routine, which needs to be called to start the lambda expression
evaluation.
In this implementation, the location of the root node in the
lambda expression graph will be used as the argument to the
\texttt{normal} routine.

\listinginput{1}{machine/machine.h}

\section{The Walker Unit}
\label{WalkApp}

The walker unit contains two commands: \texttt{init}, which
initializes the state, and \texttt{walk}, which steps through
the tree counterclockwise, i.e. the function in applications is
processed prior to the argument.

\listinginput{1}{machine/walk.c}

\section{The Storage Manager}
\label{StorageApp}

The storage manager unit consists of the \texttt{put},
\texttt{get}, and \texttt{clear} commands implementation as well
as the \texttt{reset} routine, which resets the memory into its
initial state.

\listinginput{1}{machine/alloc.c}

\section{The Copy Routine}
\label{CopyApp}

The following is the \texttt{copy} command implementation.

\listinginput{1}{machine/copy.c}

\section{The Replacement Mechanism}
\label{ReplaceApp}

The replacement mechanism is implemented here, and so is the
routine that checks if a node can be reduced.

\listinginput{1}{machine/replace.c}

\section{The Evaluator Algorithm}
\label{EvalApp}

Given below is the core algorithm of the Heap Lambda Machine.
This algorithm evaluates the lambda expression residing in the
machine memory.

\listinginput{1}{machine/normal.c}

\end{document}